\documentclass[fleqn,usenatbib]{mnras}

\usepackage{newtxtext,newtxmath}

\usepackage{tikz}
\usetikzlibrary{svg.path}
\definecolor{orcidlogocol}{HTML}{A6CE39}
\tikzset{orcidlogo/.pic={
 \fill[orcidlogocol] svg{M256,128c0,70.7-57.3,128-128,128C57.3,256,0,198.7,0,128C0,57.3,57.3,0,128,0C198.7,0,256,57.3,256,128z};
 \fill[white] svg{M86.3,186.2H70.9V79.1h15.4v48.4V186.2z}
 svg{M108.9,79.1h41.6c39.6,0,57,28.3,57,53.6c0,27.5-21.5,53.6-56.8,53.6h-41.8V79.1z M124.3,172.4h24.5c34.9,0,42.9-26.5,42.9-39.7c0-21.5-13.7-39.7-43.7-39.7h-23.7V172.4z}
 svg{M88.7,56.8c0,5.5-4.5,10.1-10.1,10.1c-5.6,0-10.1-4.6-10.1-10.1c0-5.6,4.5-10.1,10.1-10.1C84.2,46.7,88.7,51.3,88.7,56.8z};
}}
\newcommand\orcidicon[1]{\href{https://orcid.org/#1}{
\begin{tikzpicture}[yscale=-0.04,xscale=0.04,transform shape]
\pic{orcidlogo};
\end{tikzpicture}
}}

\usepackage[T1]{fontenc}

\DeclareRobustCommand{\VAN}[3]{#2}
\let\VANthebibliography\thebibliography
\def\thebibliography{\DeclareRobustCommand{\VAN}[3]{##3}\VANthebibliography}

\usepackage{graphicx}	
\usepackage{amsmath}	

\usepackage[normalem]{ulem} 

\newcommand{\sev}[2]{\bgroup\markoverwith{\textcolor{magenta}{\rule[0.5ex]{2pt}{1pt}}}\ULon{#1}\textcolor{magenta}{#2}}

\title[Accelerated phase-mixing due to a rotating bar]{Accelerated phase-mixing in the stellar halo due to a rotating bar}
\author[E. Y. Davies et al.]{
Elliot Y. Davies~\orcidicon{0000-0001-5996-4072}$^{1}$\thanks{E-mail: eyd20@cam.ac.uk},
Adam M. Dillamore~\orcidicon{0000-0003-0807-5261}$^{1}$,
Eugene Vasiliev~\orcidicon{0000-0002-5038-9267}$^{1}$ and
Vasily Belokurov~\orcidicon{0000-0002-0038-9584}$^{1}$.
\\
$^{1}$Institute of Astronomy, University of Cambridge, Madingley Road, Cambridge CB3 0HA, UK\\
}

\date{Accepted XXX. Received YYY; in original form ZZZ}

\pubyear{2023}

\begin{document}
\label{firstpage}
\pagerange{\pageref{firstpage}--\pageref{lastpage}}
\maketitle

\begin{abstract}
In a galaxy merger, the stars tidally stripped from the satellite and accreted onto the host galaxy undergo phase mixing and form finely-grained structures in the phase space. However, these fragile structures may be destroyed in the subsequent galaxy evolution, in particular, by a rotating bar that appears well after the merger is completed. In this work, we investigate the survivability of phase-space structures in the presence of a bar. We find that a bar with amplitude and pattern speed similar to those of the Milky Way would blur and destroy a substantial amount of the substructure that consists of particles with pericentre radii comparable to the bar length. While this appears to be in tension with the recent discovery of phase-space chevrons in \textit{Gaia} DR3 data, the most prominent chevrons in our simulations can still be recovered when applying the same analysis procedure as in observations. Moreover, the smoothing effect is less pronounced in the population of stars whose angular momenta have the opposite sign to the bar pattern speed.
\end{abstract}




\section{Introduction}\label{introduction}

Debris stripped from a satellite galaxy onto its host during a merger will evolve continuously in phase space, even in a static host potential where the integrals of motion remain unchanged. This phase mixing process results in the formation of fine substructure in phase space. In the case of an eccentric merger, this phase space substructure manifests as a series of chevron-like features in $(v_r,r)$ space \citep[e.g.][]{fillmore1984self,sanderson2013shells,dong20226dshells}. While initially compact in phase space, the satellite debris is stretched according to Liouville's theorem as the progenitor falls into the host potential. The merged debris then continuously winds up into finer and finer substructure in phase space as it completes radial orbits. In the case of a relatively large satellite, not all of the debris will be stripped in one go, and so several populations of wound substructure will form. This continuous evolution of substructure allows one to count the chevrons as if they were tree rings, giving a hint to the age of the merger as well as to the properties of the host potential, or the trajectory of the satellite. However, as the debris is perturbed following the phase mixing of a large merger, this substructure may not be preserved \citep[e.g.][]{davies2022Ironing}. One such perturber could be a rotating bar. 

It has long been established that the Milky Way (MW), like up to two-thirds of all disc galaxies \citep{aguerri2009population}, has a stellar bar at its centre. While originally found by gas kinematics \citep{peters1975models, binney2009understanding} and near-infrared emission \citep[][]{blitz1991direct}, evidence for the Galactic bar is now also strongly supported by stellar kinematic data \citep[e.g.][]{howard2009kinematics, shen2010milky, debattista2017separation}. 

Recently, the discovery of a high radial anisotropy population of MW halo stars in the inner stellar halo \citep[][]{belokurov2018coformation, helmi2018merger} has led to the conclusion that a dwarf galaxy, dubbed \textit{Gaia} Sausage Enceladus (GSE), merged with the MW beginning about 8 -- 11 Gyrs ago. The GSE likely dominates much of the inner stellar halo \citep[][]{iorio2019shape, lancaster2019halos} and is thought to be the most recent massive merger the MW has undergone \citep[e.g.][]{deason2013broken, evans2020how}. With a total stellar mass estimated to be about $10^8$ -- $10^9$ M$_{\odot}$, the GSE may account for up to 2/3 of the stellar mass in the MW halo \citep[][]{fattahi2019origin,dillamore2022merger,naidu2021reconstructing}. Conveniently, the debris from the last major merger may be uniquely situated to studying the kinematics of the bar, which extends no more than a few kpc from the Galactic centre \citep[e.g.][]{wegg2015structure, lucey2022constraining}, as the GSE provides a population of high eccentricity, low pericentre stars.

Observational studies into the dynamics and formation history of the MW have been revolutionised by the \textit{Gaia} mission \citep[][]{Gaia}, with its third data release (hereafter \textit{Gaia} DR3) providing tens of millions of stars with full 6d position and velocity data \citep[][]{gaiadr3}. Motivated by the apparent detection of chevron-shaped phase space structures in the \textit{Gaia} DR3 data \citep[][]{belokurov2022energy}, we investigate the behaviour and survival of these chevrons in the presence of a rotating bar. Specifically, we assume a quadrupole bar \citep[][]{dehnen2000effect, monari2016effects} that formed long enough after the most massive merger so that the deposited debris was able to phase-mix sufficiently.

The outline of this work is as follows. In Section~\ref{sec:sim_method} we briefly outline the details of our simulations.  We present the results of these simulations in Section~\ref{sec:results}, and compare them with \textit{Gaia} DR3 data in Section~\ref{sec:data}. Lastly, we summarise our results in Section~\ref{sec:summary}.

\section{Simulation Method}\label{sec:sim_method}

\subsubsection*{Initial Merger Simulation}\label{sec:mergersim}

We run an initial $N$-body merger simulation which is exactly the same as the one described in Section 4.1 of \citet[][]{davies2022Ironing}. In this simulation the host is built of a combined disk, bulge and halo potential, whereas the satellite is just made of a disk and a halo component. The merger simulation runs for a total of 5 Gyr, thereby allowing the satellite to substantially phase mix before the introduction of a bar potential. This $N$-body simulation aims to produce merged debris with properties that approximate a GSE-like merger. The $2\times10^{11}$ M$_{\odot}$ mass satellite galaxy is represented by $2\times10^5$ stellar and $8\times10^5$ dark matter particles, and the $5\times10^{11}$ M$_{\odot}$ host has twice as many particles. After evolving the $N$-body simulation for 5 Gyr with the \textsc{gyrfalcON} code \citep{dehnen2002falcon}, we create a static axisymmetric potential from the final $N$-body snapshot represented by a multipole expansion, and save the final positions and velocities of the satellite's stellar particles. We integrate the orbits of these stellar particles in the static potential of the merger remnant (host plus satellite) plus a rotating bar potential with a time-dependent amplitude, using the \textsc{Agama} code \citep{vasiliev2019agama}.

\subsubsection*{Bar Potential}

To represent the bar, we introduce a purely quadrupole potential with zero total mass \citep{dehnen2000effect} described by eqs.~1--3 in \citet{monari2016effects}.The amplitude of the bar smoothly grows from zero to a constant value $A_f$ between times $t_0$ and $t_1$ as described by eq.~4 in \citet{dehnen2000effect}, with the final amplitude equivalently expressed in terms of dimensionless parameter $\alpha$ (eq.~7 in the same paper, in which we set $R_0=8$~kpc and $v_0=230$~km/s). 

We run two grids (one positive pattern speed, one negative pattern speed) of 15 primary bar simulations each where the above values of $R_0$ and $v_0$ are always fixed. Likewise, the bar's final radius is always $R_b = 2$ kpc, the bar's time of growth is always 2 Gyr from $t_0 = 5 + 1$ Gyr to $t_1 = 5 + 3$ Gyr. We fix the bar radius to this value, in line with what is expected of galaxies around the redshift of the GSE merger \citep[e.g.][]{rosas2022evolution}. The only parameters which are varied from simulation to simulation are the bar strength $\alpha$ and the bar pattern speed $\Omega_{\rm b}$, which has units of km/s/kpc. We consider the same values of $\alpha = \{0.007, 0.010, 0.013\}$ as in \citet[][]{dehnen2000effect}, and values of $|\Omega_{\rm b}| = \{36, 38, 40, 42, 44\}$ km/s/kpc, covering a sensible range of pattern speeds determined by observations \citep[e.g.][]{sanders2019pattern, lucey2022constraining, li2022gas, leung2023measurement}. In addition to these primary bar simulations, we run two additional sets of 15 simulations with the same three values of alpha, but with $|\Omega_{\rm b}| = \{0, 5, 10, 20, 30\}$ km/s/kpc, to ensure sensible behaviour down to low pattern speed values, and to ensure any effect we see is actually dependent on the \textit{rotation} of the bar.

\section{Results}\label{sec:results}

In this section we present our findings for how a rotating bar affects the phase-mixed substructure, as a function of bar strength $\alpha$, pattern speed $\Omega_{\rm b}$ and whether the merged satellite debris is prograde or retrograde with the bar. We first explore the effect of a bar on the entire population of $2\times10^5$ merged satellite debris particles. 

\begin{figure}
    \centering
    \includegraphics[width=0.9\columnwidth]{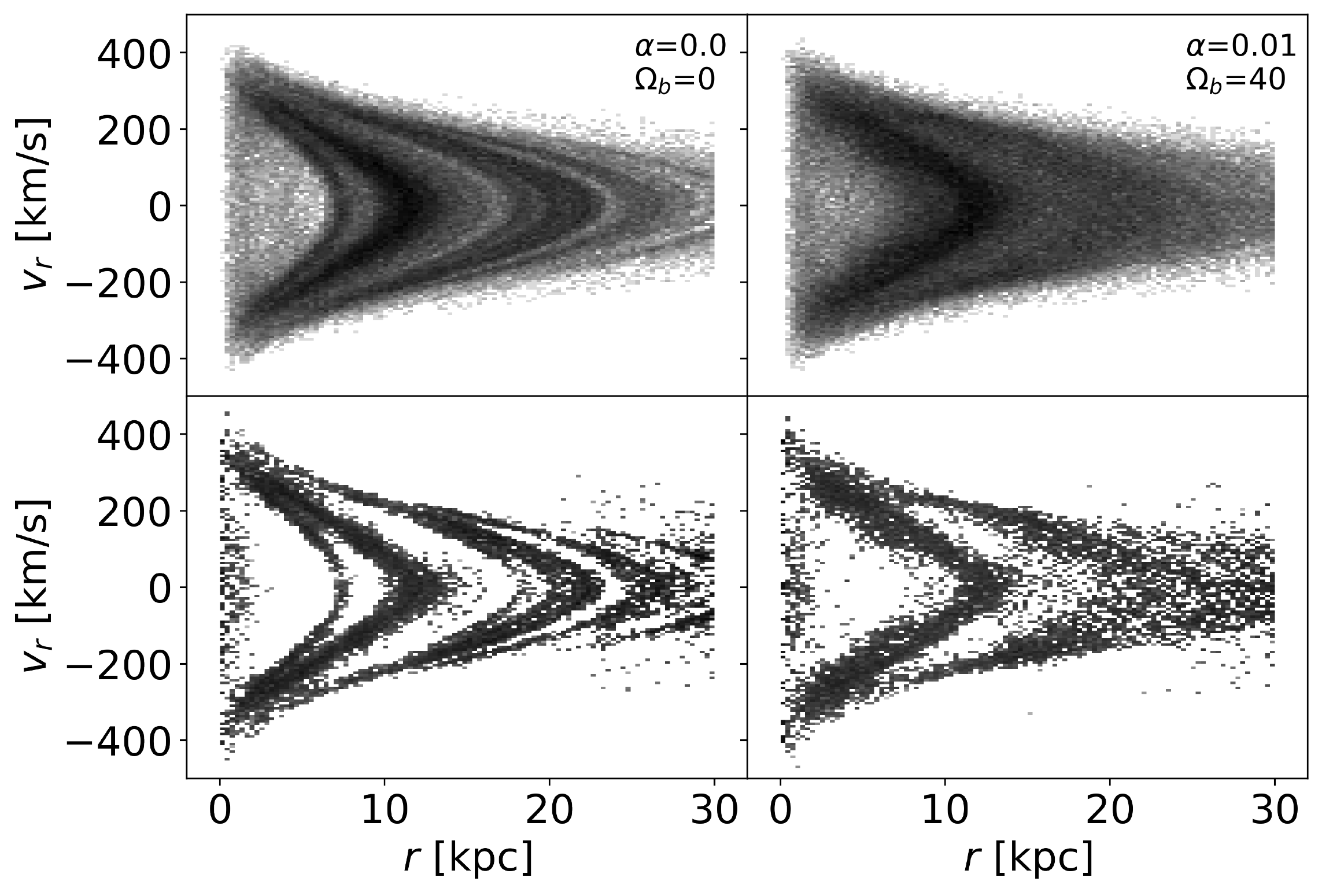}
    \caption{Final snapshot of the $(v_r,r)$ space chevrons for all $2\times10^5$ satellite debris particles in the preliminary simulation. The snapshot is taken after 5 Gyrs of the initial $N$ body, plus 5 Gyrs of test particle evolution. \textbf{Top row:} The left panel shows the evolution of the particles with no bar, whereas the right panel shows the evolution of particles with a bar of strength $\alpha = 0.01$ and pattern speed $\Omega_{\rm b} = 40$ km/s/kpc which grows for 2 Gyr, after 6 Gyr of no-bar phase mixing. \textbf{Bottom row:} The logarithm of the above density plots, column-normalised and unsharp-masked.}
    \label{fig:first_compare}
\end{figure}

After letting the merger debris evolve in the bar potential for a total of 4 Gyrs, in which the bar is growing for the first 2 Gyrs, we compare the orbital properties of the particles to a simulation in which they evolve with no bar. We first examine the appearance of the chevrons in the phase mixed satellite debris in the final snapshot of the simulation. The top row of Fig.~\ref{fig:first_compare} shows a comparison plot between a simulation with no bar (left column), in which several chevrons are clearly visible, and a simulation with a bar of strength $\alpha = 0.01$ and $\Omega_{\rm b} = 40$ km/s/kpc (right column) , in which most of the substructure disappears, except for an overdensity around 10~kpc. However, a more sophisticated analysis procedure (identical to that used in \citealt{belokurov2022energy}) reveals further details. We column-normalise the 2d density histogram and then subtract a smoothed version of it (convolved with a Gaussian filter with width of $9\times9$ pixels or $1.26 \rm \: kpc \times 60 \: \rm km/s$). The bottom row of Fig.~\ref{fig:first_compare} shows these unsharp-masked density plots, in which finer structures disappear in the presence of the bar, leaving only the largest chevron at 10~kpc and a hint of another blurry one around 20 kpc.

\begin{figure}
    \centering
    \includegraphics[width=0.90\columnwidth]{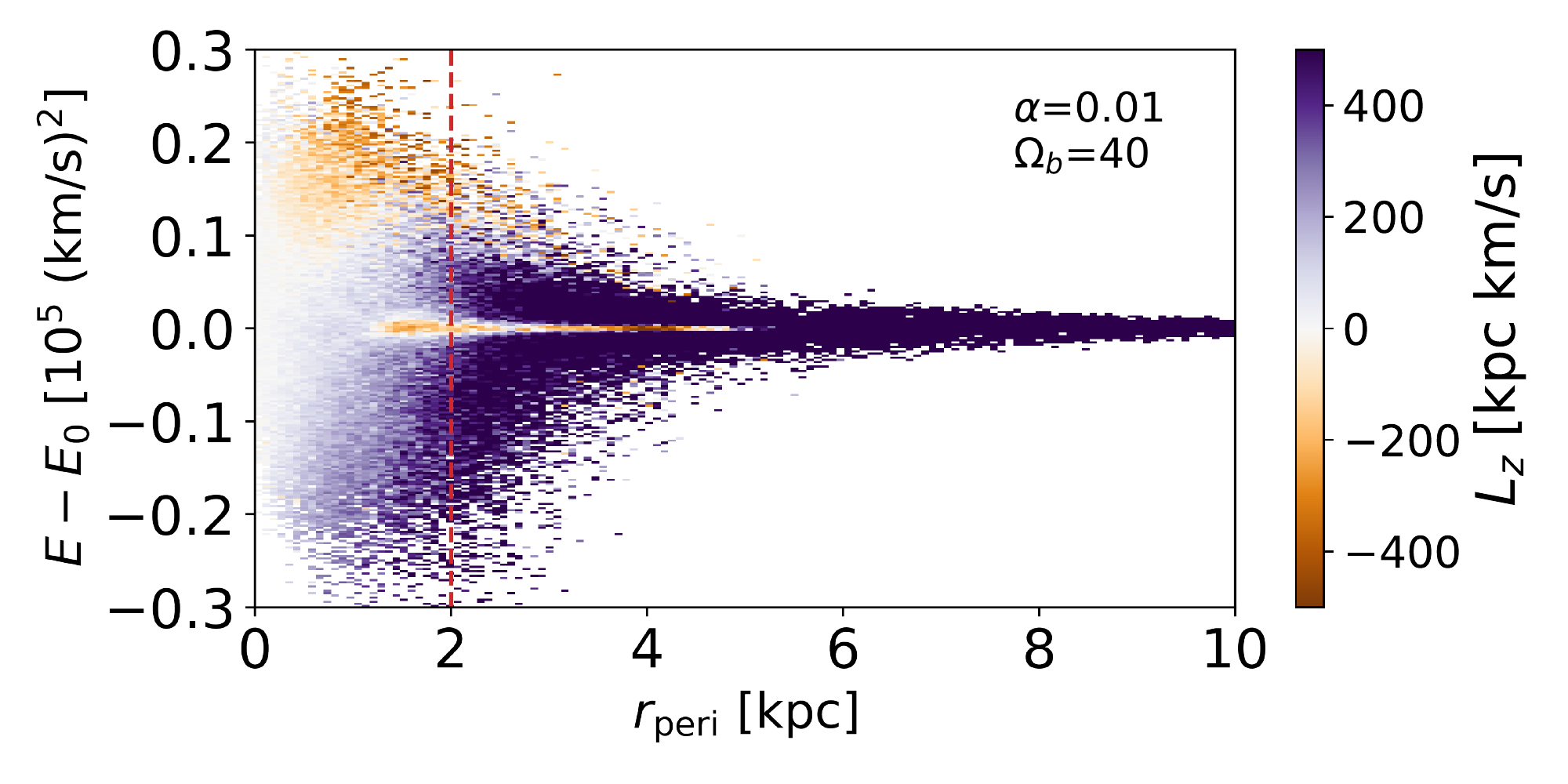}
    \caption{The energy change of each particle between the final snapshot with a bar and the final snapshot without a bar, plotted against the particles' pericentre 1 Gyr before the bar is introduced. The dashed red vertical line shows the bar radius of 2 kpc, and the plot is coloured by the median $L_z$. Evidently, particles with small pericentres, comparable to the bar's radius (2 kpc), are affected the most. Moreover, there is a preference for particles with large positive initial $L_z$ to have their energies decreases, and large negative initial $L_z$ to have their energies increased.}
    \label{fig:peris_affected}
\end{figure}

One would expect that for a particle to have its orbital properties changed by the presence of a bar, it must travel through (or close to) the bar i.e. the particles pericentre $r_{\rm peri} \lesssim r_{\rm bar}$. We confirm this assumption in Fig.~\ref{fig:peris_affected} by plotting the difference in energy for each particle in the bar simulation with energy in the no-bar simulation, $E-E_0$. It is clear that particles with lower pericentres have a higher change in energy. For reference, the bar's radius (2 kpc) is shown by the red dashed line. There is a more dramatic change as particles' pericentres get closer to the bar radius. We show this energy change for all $2\times10^5$ stellar particles as a function of pericentre for the same simulation with $\alpha = 0.01$ and $\Omega_{\rm b}$ = 40 km/s/kpc. Additionally, we colour the data by median $L_z$ to reveal a dependence of the change in energy on z-angular momentum. We note that particles whose energies increased the most were those with high negative $L_z$. Few, if any, particles with high negative $L_z$ had their energies decreased, while many particles with positive $L_z$ did. This shows a tendency for retrograde particles to increase their energy, and prograde particles to have their energies decreased. We explore this effect further in later plots.

\begin{figure}
    \centering
    \includegraphics[width=0.95\columnwidth]{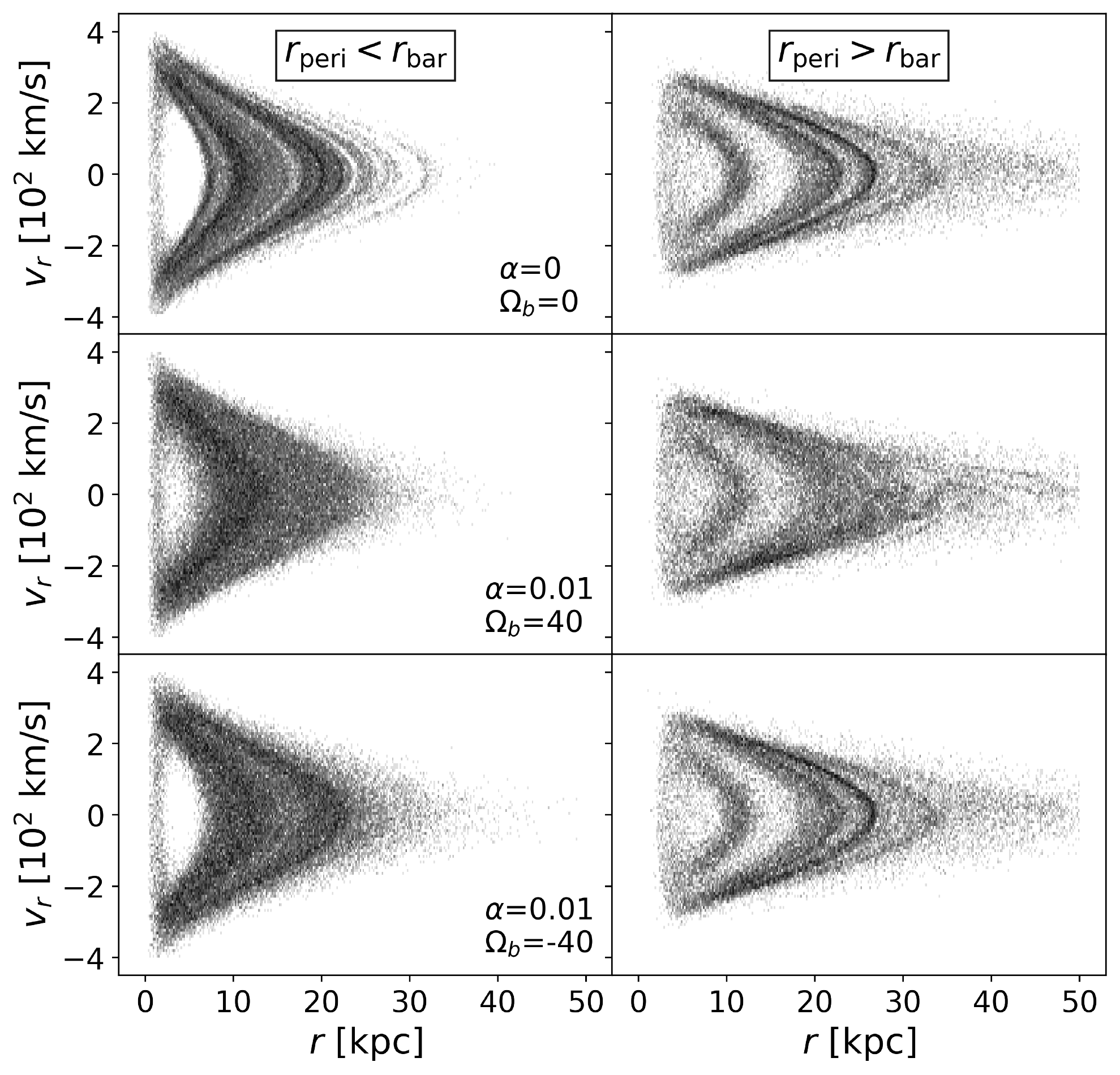}
    \caption{Final snapshot of $(v_r,r)$ space, shown for four different simulations. In all simulations shown (where a bar is present) the bar has a radius of $r_{\rm bar} = 2$ kpc, and the only particles shown are those with positive initial (1 Gyr before bar turn-on) z-angular momentum $L_z$. We cut on $L_z$ in order to show the different effect of prograde and retrograde debris. We have also split the particles into two groups; on the left we show those with $r_{\rm peri} < 2$ kpc and on the right we show those with $r_{\rm peri} > 2$ kpc. The top row is for reference, and shows the final snapshot where no bar is present i.e. $\alpha = 0.00$ and $\Omega = 0$ km/s/kpc. From top to bottom the next three rows show a non-rotating bar ($\Omega = 0$) with $\alpha = 0.01$, a rotating bar with the $\Omega = 40$ km/s/kpc and $\alpha = 0.01$, and finally a rotating bar with $\Omega = -40$ km/s/kpc and $\alpha = 0.01$.}
    \label{fig:4plot_compare}
\end{figure}

Given the distinction between particles with low pericentre (i.e. $r_{\rm peri} \lesssim r_{\rm bar}$) and with high pericentre (i.e. $r_{\rm peri} \gtrsim r_{\rm bar}$), it seems sensible to visually inspect the chevrons when binned by pericentre. In Fig.~\ref{fig:4plot_compare}, we present the $(v_r, r)$ space for 3 different example simulations, and split the particles up based on their pericentres. For this sample we choose values of $\alpha = 0.01$ and $|\Omega_{\rm b}| = 40$ km/s/kpc. Moreover, we now select particles with only $L_z>0$ ($\sim 2/3$ of the total number), and we denote the simulations with $\Omega_{\rm b}>0$ as ``prograde'' and those with $\Omega_{\rm b}<0$ as ``retrograde'', to assess the impact of varying the bar rotation with respect to the debris. Note that for an eccentric GSE-like merger, there will always be a mixture of prograde and retrograde debris.

The left column shows particles with $r_{\rm peri} < r_{\rm bar}$ and the right column shows particles with $r_{\rm peri} > r_{\rm bar}$, where $r_{\rm bar} = 2$ kpc in all cases. In both pericentres bins, the number of particles is comparable, with $n = 5.7\times10^4$ in the former and $n = 7.9\times10^4$ in the latter. From top to bottom, we show the final snapshot of a simulation with no bar, a bar with a prograde debris, and a bar with retrograde debris. We also ran a simulation with a non-rotating bar and found that the $(v_r,r)$ space was unchanged from the no bar simulation. It is clear that the inclusion of a rotating bar causes much of the substructure to be removed. However, the prograde and retrograde set-ups differ in an interesting way. While the particles with low $r_{\rm peri}$ have their substructure disturbed in both cases, albeit slightly less so in the retrograde cases, the high $r_{\rm peri}$ particles substructure is essentially unchanged in the retrograde case. After reexamination of Fig.~\ref{fig:peris_affected}, this difference between prograde and retrograde could be related to the fact that particles are more likely given an increase in energy in the former case, and a decrease in energy in the latter. To explore this in more detail, we examine the energy distribution of the debris particles.

\begin{figure}
    \centering
    \includegraphics[width=\columnwidth]{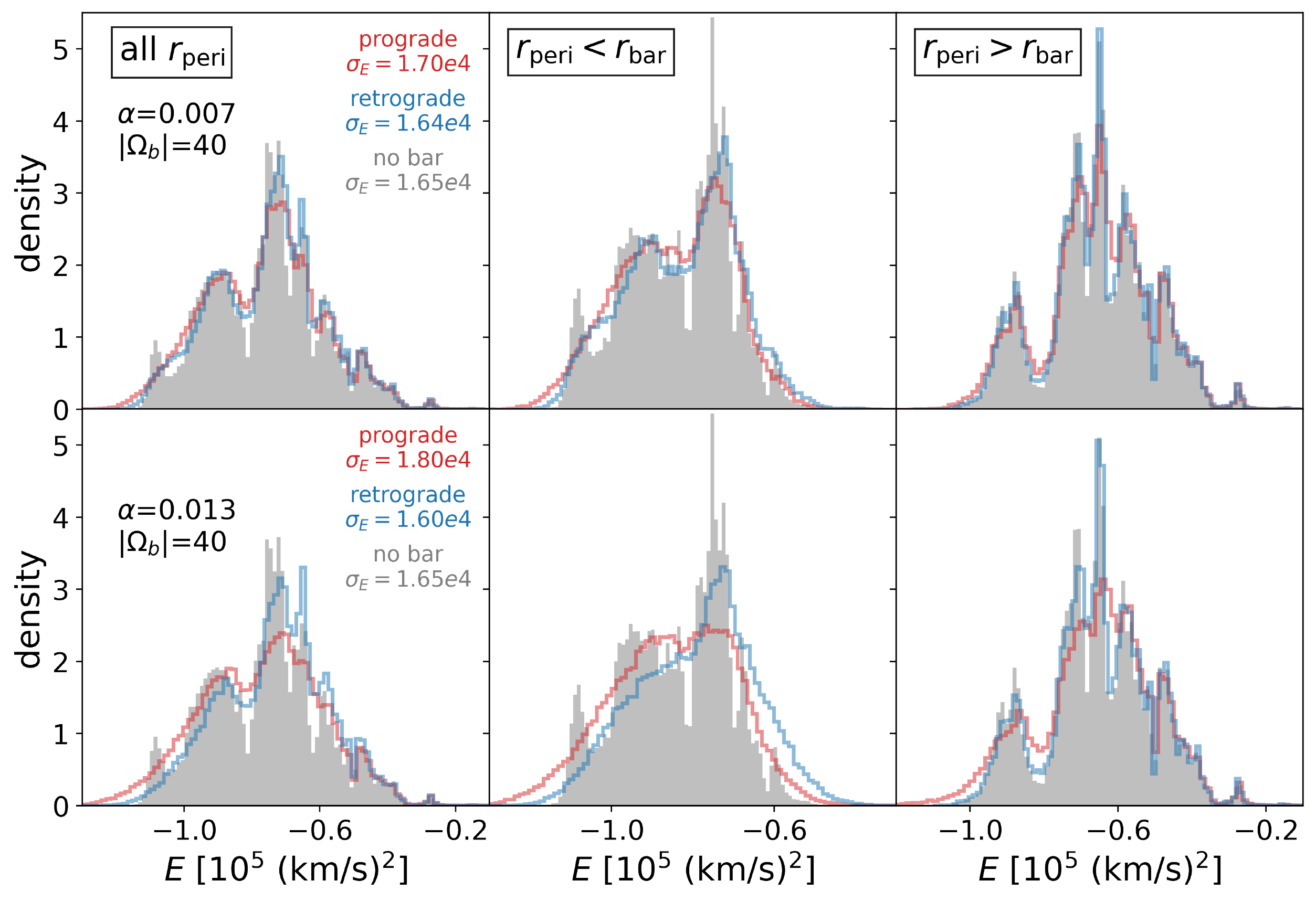}
    \caption{Histograms of energy distributions for two different bar simulations, both with $|\Omega_{\rm b}| = 40$ km/s/kpc. Again,
    we only show particles with initial $L_z > 0$. The top row shows the final snapshot energy distributions for a bar with $\alpha = 0.007$ and the bottom the distribution for $\alpha=0.013$. Each panel shows distributions for in a no-bar simulation in grey, prograde debris ($\Omega_{\rm b} = +40$) in red, and retrograde debris ($\Omega_{\rm b} = -40$) in blue. In the left column, we show particles of all pericentres, and indicate the corresponding energy spreads, $\sigma_E$. In the middle and right column we show the particles binned by pericentre within the bar ($r_{\rm peri} < 2$ kpc) radius and beyond the bar radius ($r_{\rm peri} > 2$ kpc), respectively.}
    \label{fig:compare_energies}
\end{figure}

\begin{figure}
    \centering
    \includegraphics[width=0.9\columnwidth]{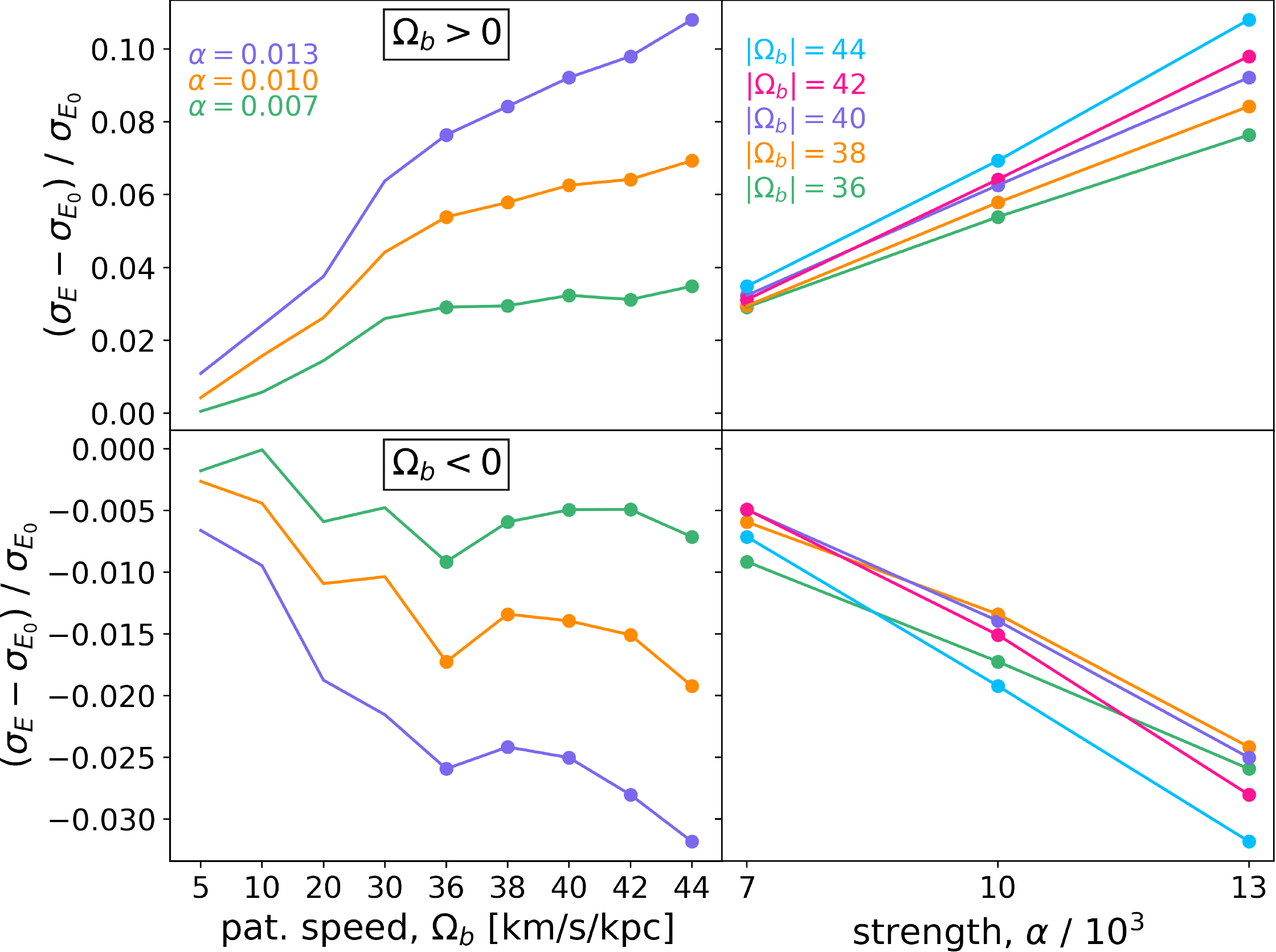}
    \caption{The change in energy standard deviation $\sigma_E$ of all $L_z > 0$ particles between simulations with a bar and simulations without a bar, plotted against bar strength $\alpha$ (right panel) and bar pattern speed $\Omega_{\rm b}$ (left panel). The top panel shows simulations where the debris is prograde ($\Omega_{\rm b} > 0$), and the bottom panel shows simulations where the debris is retrograde ($\Omega_{\rm b} < 0$).}
    \label{fig:lineplot}
\end{figure}

In Fig.~\ref{fig:compare_energies} we show the 1d energy distributions in the final snapshots of two simulations: $\alpha=0.007$ (0.013) in the top (bottom) rows. Again we split the results into two bins based on pericentre, as well as presenting them unbinned. Note that this pericentre binning also splits the populations up into a low energy and a high energy group, as expected. In all panels, we plot the final snapshot energies for the simulations with no bar (grey filled histogram), for simulations with prograde debris (red line histogram), and for simulations with retrograde debris (blue line histogram). It is clear that the energy distributions are changed more for the particles with lower pericentres than for those with higher pericentres, with much more energy substructure being retained in the high pericentre case. Moreover, we find that the energy distribution substructures are better preserved in the retrograde case. This is most easily seen in the simulation with higher $\alpha$, where the retrograde case still contains a large peak, but the prograde case has become flatter in the middle. In the leftmost column of Fig.~\ref{fig:compare_energies} we no longer bin the distribution by pericentres. Here, we show that there is an increased standard deviation of energies $\sigma_E$ for the prograde case, but a decreased $\sigma_E$ for the retrograde case, especially for the higher bar amplitude. In Fig.~\ref{fig:lineplot}, we illustrate the change in standard deviation for every simulation in the grid.  Here we see that, in all cases, retrograde debris ($\Omega_{\rm b} < 0$) causes the fractional change in standard deviation, $\delta\sigma_E~=~(\sigma_E~-~\sigma_{E_0})$~/~$\sigma_{E_0}$, to decrease, where $\sigma_{E_0}$ is the standard deviation of the energies in the simulation with no bar. Moreover, the greater value of $\alpha$ causes a greater change in $\sigma_E$. However, for the prograde case ($\Omega_{\rm b} > 0$), $\sigma_E$ is always increased, and more so at larger values of $\alpha$. The change in $\sigma_E$ appears less dependent on the value of the the pattern speed than on the bar strength. The top-left panel of Fig.~\ref{fig:lineplot} shows that $\delta\sigma_E$ does increase as $\Omega_{\rm b}$ increases, but the change is from highest to lowest $\Omega_b$ is less than the change from highest to lowest $\alpha$. Likewise, the bottom-left panel shows a reduced dependence on $\Omega_{\rm b}$ in the retrograde case, for values nearer to observational measured values of $\Omega_{\rm b}$ (plotted scatter points).

\begin{figure}
    \centering
    \includegraphics[width=0.8\columnwidth]{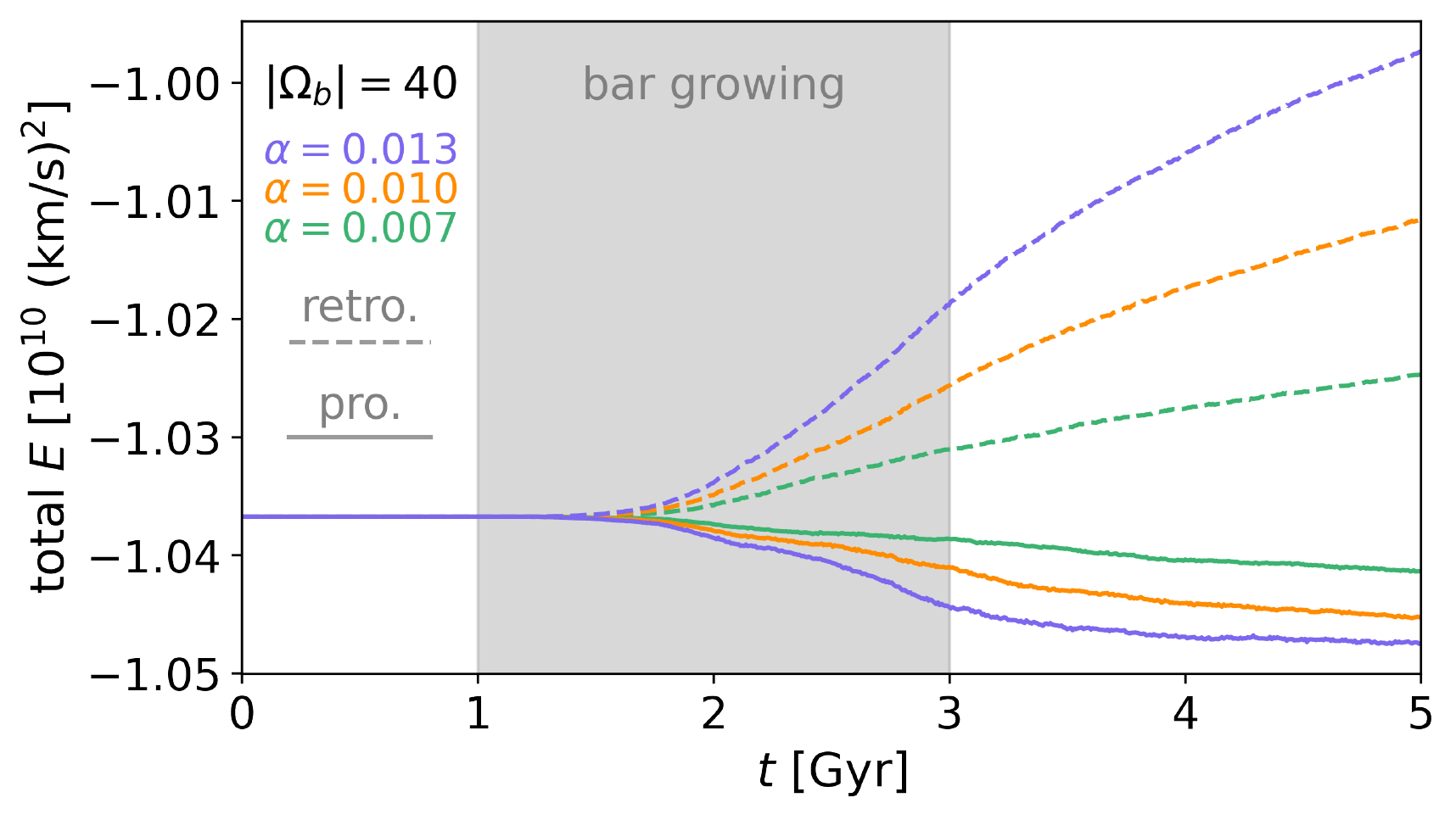}
    \caption{The total energy change of particles with $L_z > 0$ in the stellar debris as a function of time, for several simulations all with the same amplitude of pattern speed, $|\Omega_{\rm b}| = 40$ km/s/kpc. We consider both prograde $(\Omega_{\rm b} > 0)$ and retrograde $(\Omega_{\rm b} < 0)$ debris, for bar strengths of $\alpha = 0.007, 0.010$ and $0.013$. The solid lines show the energy change for prograde set-ups, while the dashed lines show the change for retrograde set-ups. The grey shaded area indicates the time in which the bar is growing from an amplitude of zero at 1 Gyr to its maximum amplitude at 3 Gyr.}
    \label{fig:energy_change_time}
\end{figure}

Lastly, we explore the change in the total energy across three simulations, this time fixing $|\Omega_{\rm b}| = 40$ km/s/kpc, as the dependence on pattern speed seems weaker than that of on bar strength. In Fig.~\ref{fig:energy_change_time} we show the total energy of all particles (with $L_z > 0$) as a function of time. The dotted lines shows retrograde simulations and the solid lines show prograde simulations. The grey shaded region indicates the time over which the bar grows from minimum to maximum amplitude. Here we see clearly that the retrograde set-up causes an increase in the total energy of the particles, while the prograde set-up causes a total decrease of the energies. Again, these effects are amplified when the bar strength is increased. 

Combining the information learned from the results above, we see that the effect of prograde debris is to decrease the energies of the low energy (low pericentre) population, while hardly changing the energies of the high-energy (high pericentre) populations. This causes the energy spread to widen, and so some substructure gets washed out. Moreover, since some particles lose energy, they fall closer into the bar and become even more impacted. To confirm this, we checked that the median Galactocentric radius of the debris particles decreased in the prograde case. However, in the retrograde case, the low energy population has an injection of energy, therefore narrowing the energy spread, since the high energy population remains essentially unchanged again. In this latter scenario, more substructure is retained since the energy distribution doesn't get diluted in the same way, and the particles are not falling further into the bar's area of influence. The chevrons that do survive appear to do so because they were initially overdense enough, and even with dilution due to changing energies there remains enough particles with similar enough energies to retain some substructure.

\section{Data Comparison}\label{sec:data}

\begin{figure}
    \centering
    \includegraphics[width=0.9\columnwidth]{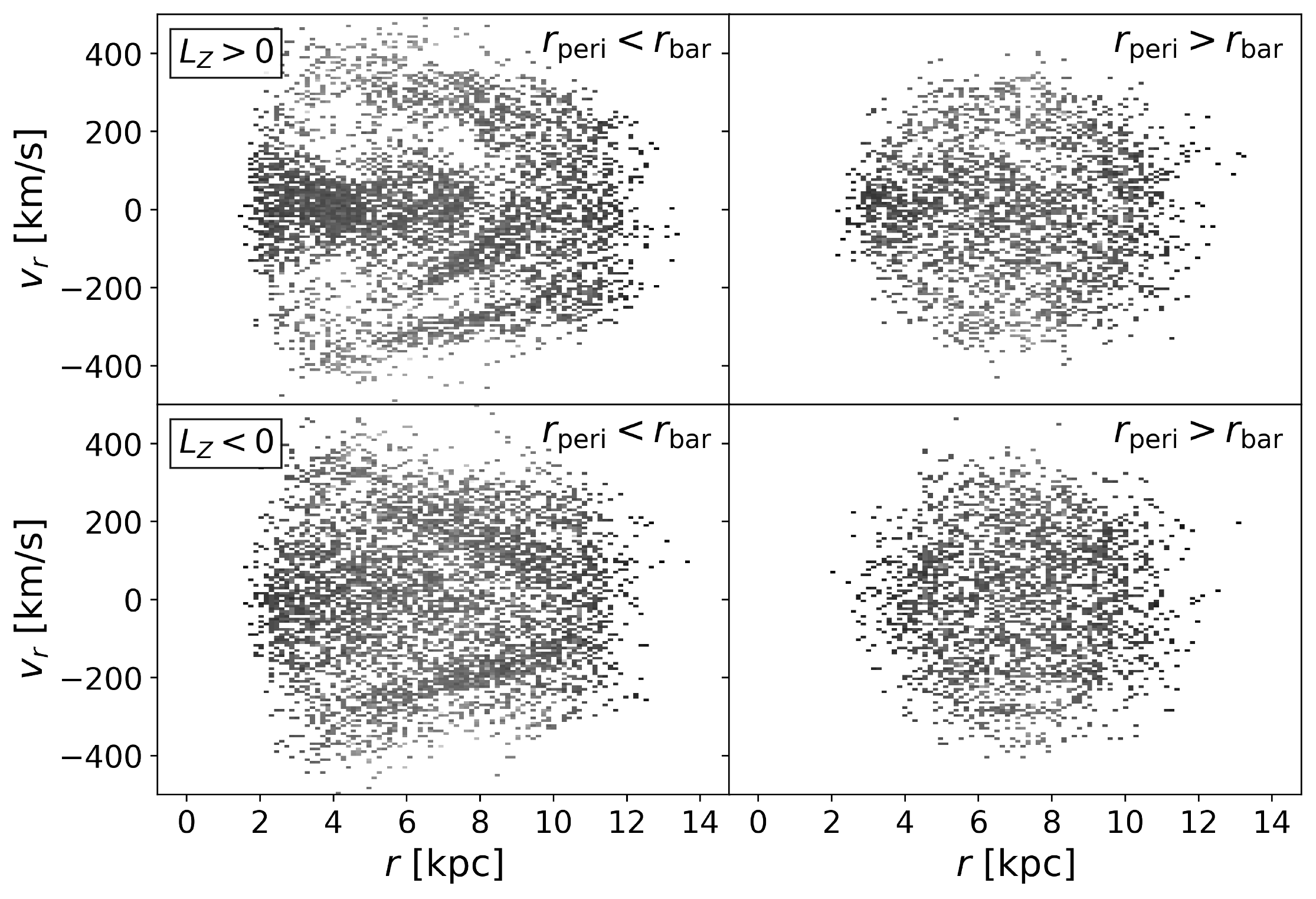}
    \caption{The local GSE $(v_r,r)$ phase space chevrons found using \textit{Gaia} DR3, binned by angular momentum and by pericentre. All panels show the logarithm of column normalised density with smooth background removed. The top row shows all stars with $L_z > 0$, whereas the bottom row shows all with $L_z < 0$. The left column shows only data with $r_{\rm peri} < 2$ kpc, and the right column shows only data with $r_{\rm peri} > 2$ kpc. Note how the chevrons are no longer visible in the right column.}
    \label{fig:data}
\end{figure}

From the results, we have learned that the bar's influence on the visual appearance of the $(v_r,r)$ chevrons is substantial for a range of strengths and pattern speeds, but more so for particles with low pericentres. With this is mind, it is important to reexamine the the chevrons discovered in \textit{Gaia} DR3. In Fig.~\ref{fig:data} we present these data, which has been cut down from the original 25 million with full 6-d information, in a similar same way as described in \citet[][]{belokurov2022energy}, ultimately providing us with about $1.5 \times 10^{5}$ halo-like stars with $|L_z| < 500$ kpc km/s. Here, we separate the data into two pericentres bins. We calculate the pericentres using the \textsc{MilkyWayPotential} from \textsc{Gala} \citep{price-whelan2017gala}. The top row shows the data for $L_z > 0$, whereas the bottom rows shows the data for $L_z < 0$. The left column shows the data with $r_{\rm peri} < 2$ kpc (quoted in the figure as $r_{\rm bar}$), and the right columns shows the with $r_{\rm peri} > 2$ kpc. From this binning, we see that the chevron-like features disappear when we consider only particles with larger pericentres. Initially, this seems somewhat in tension with the work above which suggests the bar should wipe smooth any substructure with low pericentres. However, we point back to Fig.~\ref{fig:first_compare}, where the bottom row has been processed in the same was the \textit{Gaia} DR3 data. This figure illustrates that some phase-mixed substructure actually remains recoverable even in the case of a rotating bar. However, the existence of the chevrons in the data \textit{only} within influence of the bar points to the possibility that they are not the result of phase mixing and instead are formed via resonance effects of the bar. We explore this bar shepherding of the halo sub-structure further in an upcoming companion paper by \citet[][]{dillamore2023stellar}.

\section{Summary}\label{sec:summary}

In this work we explore the impact of a rotating bar on the phase-space substructure that results from the phase mixing of debris from a large $N$-body merger. We allow the merger debris to phase mix for 5 Gyr in a live $N$-body simulation, and then switch to a test-particle integration of debris orbits in an axisymmetric potential of the merger remnant, into which a bar component is gradually added over the course of 2 Gyr. We run two sets of 30 bar test particle simulations, where all parameters of the bar are fixed except for the values of the bar strength $\alpha = \{0.007, 0.010, 0.013\}$ and the values of the pattern speed $\Omega_{\rm b} = \{0, 5, 10, 20, 30, 36, 38, 40, 42, 44\}$ km/s/kpc. In future work we intend to study the effect of varying the bar length.

We show how the bar impacts the visual presentation of the $(v_r,r)$ phase space substructure, as well as the distribution of energies. We bin the results by pericentre, separating particles with $r_{\rm peri} < r_{\rm bar}$ from those with $r_{\rm peri} > r_{\rm bar}$. Additionally, we investigate the difference between prograde debris and retrograde debris. Lastly, we compare our results to the data discovered in \textit{Gaia} DR3 by \citep[][]{belokurov2022energy}. Our findings can be summarised as follows:

\begin{enumerate}
    \item Particles with $r_{\rm peri} < r_{\rm bar}$ have their orbital properties changed more than those with $r_{\rm peri} > r_{\rm bar}$.
    \item Satellite debris particles that are retrograde (prograde) with the bar have their total energy increased (decreased) and their energy spread decreased (increased).
    \item In all simulations with bar pattern speed $\Omega_{\rm b}$ comparable to that of the Milky Way, the chevrons in the $(v_r,r)$ phase space are substantially blurred. However, some of the substructure is recoverable when an unsharp-masking filter is applied, as with the \textit{Gaia} DR3 data.
    \item The visual appearance of the $(v_r,r)$ substructure is much less impacted in the case of retrograde debris than in the case of prograde debris. 
    \item The chevrons found in the \textit{Gaia} DR3 data seem to consist entirely of stars with low pericentre, which may  suggests that they are the result of bar related phenomena, such as resonant interactions.
\end{enumerate}

\appendix

\section*{Acknowledgements}

EYD thanks the Science and Technology Facilities Council (STFC)
for a PhD studentship (UKRI grant number 2605433). AMD thanks STFC for a PhD studentship (UKRI grant number 2604986). The authors thank the anonymous referee for helpful comments.

\section*{Data Availability}

The simulations and analysis in this project can be reproduced with publicly available software. We also make use of publicly available {\it Gaia} DR3 data.



\bibliographystyle{mnras}
\bibliography{bar_impact} 




\appendix


\bsp	
\label{lastpage}
\end{document}